\documentclass[laps,pra,twocolumn,floatfix]{revtex4}

\usepackage[mathscr]{eucal}
\usepackage{amsfonts}
\usepackage{hyperref}
\usepackage{amsmath,amssymb}
\usepackage{color}
\usepackage{graphics}
\usepackage {epsfig}
\usepackage{psfrag}
\usepackage{graphicx}

\begin{document}

\title{High-efficiency cluster-state generation with atomic ensembles
via the dipole-blockade mechanism}
\author{Marcin Zwierz}\email{php07mz@sheffield.ac.uk}
\author{Pieter Kok}
\affiliation{Department of Physics and Astronomy, University of
Sheffield, Hounsfield Road, Sheffield, S3 7RH, UK}

\date{\today}

\begin{abstract}
We demonstrate theoretically a scheme for cluster state generation, based on atomic
ensembles and the dipole blockade mechanism. In the protocol, atomic ensembles serve as single qubit systems.
Therefore, we review single-qubit operations on qubit defined as collective states of atomic ensemble.
Our entangling protocol requires nearly identical single-photon sources, one ultra-cold ensemble per physical qubit, and regular
photodetectors. The general entangling procedure is presented, as well as a
procedure that generates \emph{in a single step} $Q$-qubit GHZ states with success probability
$p_{success}\sim\eta^{Q/2}$, where $\eta$ is the combined detection and source
efficiency. This is significantly more efficient than any known robust
probabilistic entangling operation. GHZ states form
the basic building block for universal cluster states, a resource for the one-way quantum computer.
\end{abstract}

\maketitle

\section{Introduction}
The construction of a quantum computer is an important goal of
modern physics. One of the prominent approaches to the physical implementation of a quantum computer and quantum computation is linear quantum optics \cite{knill} \cite{koklinear}.
Photons are perfect carriers of quantum information and can therefore be utilized in quantum communication \cite{nielsen}.
The drawback of photonic systems for quantum computation is the fact that there is no direct interaction between photons.
Moreover, all linear optical gates that process quantum information are probabilistic. These problems impose a requirement for a medium that
would facilitate an interaction between photons, and store photonic qubits. Hence, a concept of optical quantum memory realized in atomic vapour
(atomic ensemble) was introduced.
Atomic ensembles consist of several hundreds of the same species of atoms kept in room temperature or cooled to $\mu$K temperature.
A large number of atoms increases the coupling strength of an interaction between light and matter, and therefore allows us to coherently manipulate
the quantum state of the ensemble with light and vice versa.
Moreover, a large number of atoms helps to suppress the negative impact of
decoherence on an information stored in atomic ensemble \cite{fleisch,lukin,duan,barrett,hammerer}.\\
Initially, atomic vapours were used as a fast quantum memory.
However, it is also possible to define a ``good'' qubit in an atomic ensemble, and the question
remains how to implement the entangling operations between the qubits
that enable universal quantum computation. It suffices to create a large
entangled multi-qubit resource ---the cluster state--- after which the
entire computation proceeds via single-qubit measurements
\cite{rauss,hein}. Cluster states are large arrays of isolated qubits connected (entangled) via $CZ$ operations. The cluster states are a scalable resource and can be built up
with probabilistic entangling operations with $p_{success} > 0$ \cite{kok}. When the success probability of entangling operation is low, a very large overhead in optical elements is required. Moreover, finite coherence times of the qubits limit practical use of the cluster states. Hence, it is extremely important to build them up in the efficient way. Here, we show how to efficiently create these cluster states
using single photons interacting with an atomic ensembles via the dipole blockade mechanism. The protocol requires identical
single-photon sources, one ensemble per physical qubit placed in the arms of a Mach--Zehnder interferometer, and regular
photodetectors. We present a general entangling procedure, as well as a
procedure that generates $Q$-qubit GHZ states with success probability
$p_{success}\sim\eta^{Q/2}$, where $\eta$ is the combined detection and source
efficiency. This is significantly more efficient than any known robust
probabilistic entangling operation \cite{lim,kok}. The GHZ states are locally equivalent to the cluster state and form
the basic building block for universal cluster states. Our protocol significantly reduces an overhead in optical elements and leads to better quantum computing prospects.

The paper is organized as follows. In Sec. II, we introduce a notion of the Rydberg state and the dipole blockade mechanism.
In Sec. III, we review the concept of an atomic ensemble as single qubit system and single-qubit operations
in atomic ensembles.
In Sec. IV, we review the one-way model of quantum computation realized on the cluster states.
We also give a description of a new entangling operation and consider its usefulness for generation of the GHZ and cluster states.
In Sec. V, we consider all major errors and decoherence mechanisms that enter the entangling procedure.
Moreover, we describe how our ideas can be implemented experimentally.

\section{Rydberg state and dipole blockade mechanism}
The Rydberg state is a state of an alkali atom characterized by a high principal quantum number $n$ \cite{gallagher}.
Rydberg atoms possess a number of interesting properties. To begin, Rydberg atoms are very large compared to normal atoms.
The radius of a Rydberg atom scales as $n^2 a_{0}$, where $a_{0}$ is the Bohr radius. Therefore, the valence electron is very weakly bound to the nucleus. The binding energy of a Rydberg state is given by
\begin{equation}
E = -\frac{R}{(n-\delta)^2}=-\frac{R}{n^{*^{2}}},
\end{equation}
where $R$ is the Rydberg constant, $n^{*}$ is the effective quantum number, and $\delta$ is the quantum defect which corrects for the deviation
from a case of the hydrogen atom \cite{li}.
\begin{figure}[!h]
\begin{center}
\includegraphics[scale=0.60]{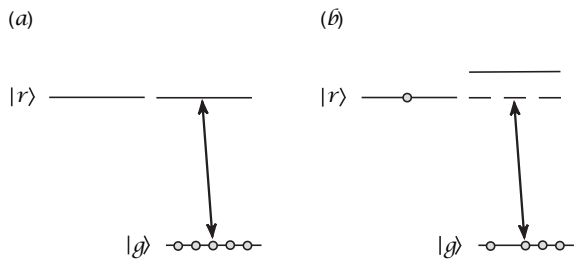}
\end{center}
\caption{The dipole blockade mechanism. A laser pulse couples ground state $|g \rangle$ and Rydberg state $|r \rangle$.
(a) One of the atoms is excited to the Rydberg state $|r \rangle$. (b) Presence of a single atom in the Rydberg state
$|r \rangle$ shifts energy levels and blocks further excitations. \label{dipole}}
\end{figure}
The Rydberg states have a very long lifetime, which scales as $\tau_{0} n^5$ where $\tau_{0}$ is the typical
lower level lifetime of around $\sim$10 ns. Hence, Rydberg states possess lifetimes of order of ms and even longer.
One of the most important properties of the Rydberg atoms is their sensitivity to external electric fields.
The Rydberg energy levels are easily perturbed by modest electric fields. Higher electric fields can even ionize
Rydberg atoms. Commonly, ionization is used as one of the detection methods. This sensitivity to electric fields is the source of a phenomenon called the \emph{dipole blockade mechanism}.
Atoms in Rydberg states have large dimensions and large dipole moments, resulting in a strong dipole--dipole interaction \cite{gerry}.
The dipole blockade mechanism was observed experimentally in small clouds of alkali atoms, such as Rubidium in a vapour cell \cite{johnson,van}.
This mechanism prevents populating states of an atomic ensembles with two or
more atoms excited to the Rydberg level \cite{lukin}.
A single atom in a micron-sized atomic ensemble excited to a Rydberg state with a narrowband laser can inhibit excitation of the other atoms in
the sample if the long range Rydberg-Rydberg interactions are much larger than a linewidth of the Rydberg state.

The physics of the dipole blockade is shown in [Fig. \ref{dipole}]. An optical pulse resonant
with a transition to the Rydberg state  $|r \rangle$ will create a Rydberg atom with a
very large dipole moment [Fig. \ref{dipole}a]. When the atoms in the ensemble are
sufficiently close, the long range Rydberg-Rydberg interactions (dipole interactions) between the Rydberg atom and
the other atoms will cause a shift in the Rydberg transition energy of
the other atoms. Therefore, the optical pulse becomes off-resonant with
the other atoms, and the ensemble is transparent to the pulse.
Under dipole blockade conditions, the mesoscopic vapour behaves as one superatom with a two-level structure. A single excitation is shared by all atoms in a sample and Rabi oscillations can be observed. Effectiveness of the blockade depends on an average strength of the interaction between atoms in the ensemble.

Depending on the separation between atoms, long range Rydberg-Rydberg interactions have different type. The usual van der Waals interaction of types
$C_{5}/R^{5}$ or $C_{6}/R^{6}$ can be resonantly enhanced by F\"{o}rster processes to the $C_{3}/R^{3}$ long range interaction.
In the absence of an external electric field, the Rydberg-Rydberg interactions are of the van der Waals type $C_{5}/R^{5}$ or $C_{6}/R^{6}$ \cite{walker}.
In a static electric field, a Rydberg atom possesses a large permanent dipole moment $p$, which scales as $\sim q a_{0} n^2$ with
$q$ the electron charge, which leads to a much stronger and longer $C_{3}/R^{3}$ interaction.
A pair of Rydberg atoms $i$ and $j$ interact with each other via dipole-dipole potential $V_{dd}$,
\begin{equation}
V_{dd} = \frac{\textbf{p}_{i}\textbf{p}_{j} - 3(\textbf{p}_{i}\cdot\textbf{e}_{ij})(\textbf{p}_{j}\cdot\textbf{e}_{ij})}{4 \pi \epsilon_{0} |\textbf{r}_{i} - \textbf{r}_{j}|^3} = \frac{p^2}{4 \pi \epsilon_{0} R^3}(1 - 3 \cos^2\theta),
\end{equation}
where $\textbf{e}_{ij}$ is a unit vector along the interatomic direction, $\theta$ is the angle between the interatomic separation $R = |\textbf{R}| = |\textbf{r}_{i} - \textbf{r}_{j}|$ and the electric field
$\textbf{z}$ direction. In general, the interaction between Rydberg atoms is very strong. However, for some angles $V_{dd}$ vanishes which is
undesirable for dipole blockade purpose \cite{walker}.
Fortunately, there is another method to induce a strong, isotropic interaction between Rydberg atoms, comparable to $V_{dd}$. The resonant
collisional process (F\"{o}rster process) transfers energy between two atoms through the dipole-dipole interaction with strength
$\sim \rho_{1} \rho_{2}/R^3$, where $\rho_{1}$ and $\rho_{2}$ are the dipole matrix elements between initial and final energy states of the
interacting atoms \cite{stoneman}. Therefore, the usual van der Waals interaction can be resonantly enhanced by F\"{o}rster processes such as
\begin{equation}
nl + nl \rightarrow n'l' + n''l''
\end{equation}
when the $nl + nl$ states are degenerated in energy with the $n'l' + n''l''$ states. The F\"{o}rster process induces an interaction potential of the form
\begin{equation}
V_{\pm}(R) = \frac{\delta}{2} \pm \sqrt{\frac{4 U_{3}(R)^2}{3} + \frac{\delta^2}{4}},
\end{equation}
where
\begin{equation}
U_{3}(R) = q^2 \langle nl || r || n'l' \rangle \langle nl || r || n''l'' \rangle / R^3,
\end{equation}
with $\delta = E(n'l') + E(n''l'') - 2E(nl)$ is the F\"{o}rster energy defect. There is no angular dependence for the potential
$V_{\pm}(R)$ so an interaction is isotropic. For perfect F\"{o}rster degeneracy ($\delta = 0$) $V_{+}(R)$ would be of similar strength
and range to $V_{dd}$ \cite{walker}. Although at the large separations, a non-zero F\"{o}rster energy defect reduces long-range interaction
between the atoms to be van der Waals $C_{6}/R^{6}$ type. However, if the F\"{o}rster energy defects are smaller than the fine-structure splitting,
then strong $C_{3}/R^{3}$ interaction occurs at longer range as well.

Although F\"{o}rster processes are very promising as a method to induce very long-range $C_{3}/R^{3}$ interactions, there are some selection
rules that need to be fulfilled for obtaining high fidelity dipole blockade. Only for $l' = l'' = l+1$ there are no so-called F\"{o}rster zero states with
$C_{3} = 0$ \cite{walker}. Therefore, a fidelity of the dipole blockade mechanism may be reduced due to weakest interactions between
degenerate Rydberg states. In the case of the F\"{o}rster zero states, a strength of the interaction between Rydberg atoms is not enhanced
and reduces to the usual van der Waals long-range type. Hence, for attaining a strong dipole blockade, resonances need to be tuned with an
electric field \cite{walker}. The other possibility for attaining strong dipole blockade is to rely on the van der Waals interaction
which at smaller distances, less than 5 $\mu$m, is large enough to mix the fine-structure levels together, so the interaction is of the $V_{dd}$ type \cite{walkerbloc}.

The dipole blockade mechanism has been proposed as a method to entangle large numbers of atoms \cite{lukin}. The exact strength of the
dipole blockade is not important as long as it is greater than the linewidth of a Rydberg state. Hence, the atoms can be at random
distances $R$ from each other \cite{walker}. Moreover, since states with two or more atoms in the Rydberg state are never populated, the atoms avoid mechanical
interactions between each other. Therefore, the atoms avoid heating and the internal states of the atoms are decoupled from the
atomic motion \cite{lukin}.

The range and quality of the dipole interaction has been studied
extensively: Walker and Saffman analyzed the primary errors that enter
the blockade process \cite{walkerbloc,van}. For Rubidium atoms with
principal quantum number $n=70$, the blockade energy shift is
approximately 1 MHz. Hence, a strong and reliable blockade is possible
for two atoms with separation up to $\sim$ 10 $\mu$m \cite{walkerbloc}.
Moreover, decoherence associated with spontaneous emission from
long-lived Rydberg states can be quite low ($\sim$ 1 ms). The dipole
blockade mechanism can be used to build fast quantum gates, i.e., a two
qubit phase gate \cite{jaksch,lukingate,brion}. The long-range
dipole-dipole interaction between atoms can be employed to realize a
universal phase gate between pairs of single-photon pulses
\cite{friedler,mohapatra,petrosyan}. Most importantly, the ideas based on the dipole blockade mechanism are experimentally feasible.

The single quantum sensitivity suggests that the dipole blockade can be used to create cluster states:
The blockade mechanism can be used in a heralding type of entangling operations and render them to be nearly deterministic. Before introducing the
entangling protocol, let us review a scheme for implementing single-qubit operations on the qubit defined in atomic ensemble and analyze it in detail.

\begin{figure}
\includegraphics[scale=0.60]{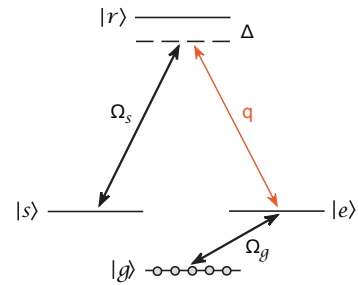}
\caption{Relevant atomic level structure with allowed atomic
transitions. States $|g \rangle$, $|e \rangle$, and $|s \rangle$ can be
realized by the electronic low-lying, long lived states of alkali atoms. The
transition between states $|g \rangle$ and $|s \rangle$ is always
dipole-forbidden. The state $|g \rangle$ is coupled to the state $|e \rangle$
through a classical field $\Omega_{g}$. A second
classical field $\Omega_{s}$ is applied to the transition between the highly
excited Rydberg level $| r\rangle$ and the state $| s \rangle$ (it may possibly be a two-photon process). States $|e
\rangle$ and $|r\rangle$ are coupled via a quantum field. In general, $\Delta$ is a small detuning. \label{levels}}
\end{figure}

\section{Atomic ensemble as single qubit system -- Single-qubit gates}
\begin{figure}
\includegraphics[scale=0.50]{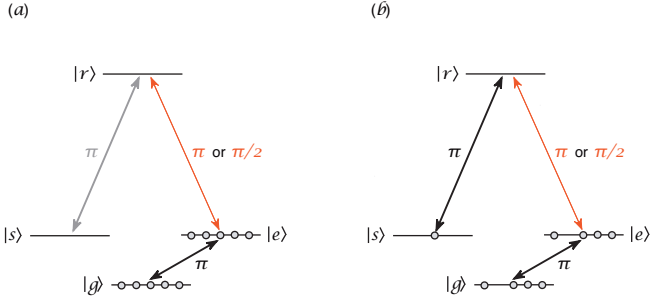}
\caption{The bit flip operation ($X$) and the Hadamard gate ($H$). (a) Rotation $| 0 \rangle_{L} \rightarrow | 1
\rangle_{L}$ or $| 0 \rangle_{L} \rightarrow 1/\sqrt{2}(| 0 \rangle_{L} + | 1 \rangle_{L}) $ (b) rotation $| 1 \rangle_{L} \rightarrow | 0 \rangle_{L}$ or $| 1 \rangle_{L} \rightarrow 1/\sqrt{2}(| 0 \rangle_{L} - | 1 \rangle_{L}) $. See the text for an explanation. \label{X&H}}
\end{figure}

A qubit may be represented by a micron-sized atomic ensemble, cooled to $\mu$K temperatures by the far off-resonant optical trap or magneto-optical trap (MOT).
The $N$ atoms at positions $r_{j}$ in an ensemble have three lower, long lived energy states $|g
\rangle$, $|e \rangle$, and $|s \rangle$ (see Fig.~\ref{levels}). The qubit
states in a mesoscopic ensemble are collective states given by
\begin{eqnarray}
|0\rangle_{L} &\equiv& |g\rangle = |g_{1},g_{2}, \ldots ,g_{N}\rangle\, , \\
|1\rangle_{L} &\equiv& |s\rangle = \frac{1}{\sqrt{N}} \sum_{j=1}^{N}
e^{i \textbf{k} \cdot \textbf{r}_{j}} |g_{1},g_{2}, \ldots ,s_{j}, \ldots ,g_{N}\rangle.
\end{eqnarray}
The states $|e \rangle$ and $|r\rangle$ participate in the
interaction part of the scheme. Levels $|g \rangle$ and $|s \rangle$
play the role of a storage states. In the case of the qubit states defined as collective states of mesoscopic ensemble, single-qubit
operations are more complex than in a case of a qubit realized on a single atom. The simplest approach to this problem is to
realize single-qubit rotations by means of classical optical pulses and the dipole blockade
mechanism. In a paper by Brion, M{\o}lmer, and Saffman \cite{brion}, the single-qubit rotations are performed by means of three laser pulses (see Fig.~\ref{X&H}).
First, a $\pi$-pulse transfers the population from $| s \rangle$ to $|r\rangle$, then a
coherent coupling of states with zero and one Rydberg excited atom is applied
for appropriate amount of time and finally, a $\pi$-pulse transfers the population from $|r\rangle$ to $| s \rangle$ (the reader may notice that two additional $\pi$-pulses are depicted in Fig.~\ref{X&H} that transfer population from the storage level $|g\rangle$ to $|e\rangle$ and back again). Therefore, in the case of a bit flip operation ($X$) the coherent coupling is just a $\pi$-pulse with a real Rabi frequency,
and the Hadamard gate ($H$) can be performed by a $\pi/2$-pulse on the same transition.
An arbitrary phase gate $\Phi(\phi) = \exp(-i\phi Z/2)$ is
realized by a detuned optical pulse applied to the transition between
$|s\rangle$ and an auxiliary level $|a\rangle$ (not shown in Fig.~\ref{X&H}).
The gates $\Phi(\phi)$, $X$, and $H$ generate all
single-qubit operations.  The readout of a qubit is based on the resonance fluorescence and again requires an auxiliary level $|a\rangle$.
An optical laser drives a transition between $|s\rangle$ and $|a\rangle$ producing a large number of fluorescence photons.
If the measurement gives no fluorescence photons, the
qubit is in  $|0\rangle_{L}$. Otherwise, a state of the qubit is projected into $| 1 \rangle_{L}$.

The scheme for implementing single-qubit operation relies heavily on the dipole blockade mechanism.
We analyze the above scheme for the case of a bit flip operation $X$. Therefore, we need to consider carefully the evolution of the system under a $\pi$-pulse applied to the transition between $|e\rangle$ and $|r\rangle$.

\begin{figure}
\includegraphics[scale=0.60]{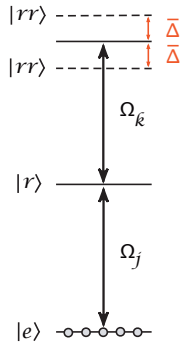}
\caption{The collective states of a mesoscopic atomic ensemble. $|e \rangle$ is the collective low-lying state, $|r \rangle$ is the singly excited state,
$|rr \rangle$ is the doubly excited state, and $\bar{\Delta}$ is the mean dipole shift.  \label{collective}}
\end{figure}

In general, the interaction of atoms with an optical laser pulse, within the dipole approximation and in the rotating frame approximation, is
governed by interaction Hamiltonian $\hat{H}_{int}$
\begin{eqnarray}
\hat{H}_{int} &=& -i \hbar \sum_{j=1}^{N} \Omega_{j} \ \sigma^{j}_{re} \ \mbox{exp}[i(\omega_{re}-\omega)t] \\
&& -i \hbar \sum_{j,k>j}^{N} \Omega_{k} \ \sigma^{jk}_{rr} \ \mbox{exp}[i(\omega_{re}-\omega)t] + \mbox{H.c.} ,
\end{eqnarray}
where $\Omega_{j} = \Omega e^{i \textbf{k} \cdot \textbf{r}_{j}}$ is the Rabi frequency, $\omega = k c$ is the frequency of an optical laser pulse,
$\sigma^{j}_{re} = |r_{j} \rangle \langle e|$
and $\sigma^{jk}_{rr} = |r_{j} r_{k} \rangle \langle r_{j}|$ are the atomic transition operators (see Fig.~\ref{collective}) \cite{saffman}.
The first transition operator $\sigma^{j}_{re}$ corresponds to the transition between the collective state $|e \rangle$
and the singly excited state
\begin{equation}
|r \rangle = 1/\sqrt{N} \sum_{j=1}^{N} e^{i \textbf{k} \cdot \textbf{r}_{j}} | r_{j} \rangle.
\end{equation}
The second one corresponds to the transition between the singly excited state $|r \rangle$ and the doubly excited state
\begin{equation}
|rr \rangle = \sqrt{2/N(N-1)} \sum_{j,k>j}^{N} e^{i (\textbf{k} \cdot \textbf{r}_{j} + \textbf{k} \cdot \textbf{r}_{k})} | r_{j} r_{k} \rangle.
\end{equation}
We assume that the optical laser pulse is resonant with a transition between $|e\rangle$
and $|r\rangle$ ($\omega_{re}-\omega = 0$). Then, the dipole interaction between two Rydberg atoms is given by
\begin{equation}
\hat{V}_{dd} = \hbar \sum_{j,k>j}^{N} \Delta_{jk} |r_{j} r_{k} \rangle \langle r_{j} r_{k}|,
\end{equation}
where $\Delta_{jk} = \frac{C_{6}}{|\textbf{r}_{j} - \textbf{r}_{k}|^6} $ is the dipole shift of the weakest van der Waals type.
Hence, the coupling of levels $|e \rangle$ and $|r \rangle$ is described by the Hamiltonian $\hat{H} = \hat{H}_{int} + \hat{V}_{dd}$.
The state vector of an atomic ensemble is given by
\begin{equation}
|\psi(t) \rangle = c_{g} |g \rangle + \sum_{j=1}^{N} c_{j} e^{i \textbf{k} \cdot \textbf{r}_{j}} |r_{j} \rangle + \sum_{j,k>j}^{N} c_{jk} e^{i (\textbf{k} \cdot \textbf{r}_{j} + \textbf{k} \cdot \textbf{r}_{k})} |r_{j} r_{k} \rangle.
\end{equation}
In the limit where the dipole shift is much larger than the Rabi frequency of an optical laser pulse $\Delta_{jk} \gg \Omega_{j}$,
the Schr\"{o}dinger equation for amplitudes of the state vector gives
\begin{eqnarray}
\dot{c}_{g} &=& \sqrt{N} \Omega c_{r}, \\
\dot{c}_{r} &=& -\sqrt{N} \Omega c_{g} + \frac{\Omega}{\sqrt{N}} \sum_{j,k>j}^{N} c_{jk}, \\
\sum_{j,k>j}^{N} \dot{c}_{jk} &=& - \sum_{j,k>j}^{N} \Omega c_{j} - i \sum_{j,k>j}^{N} c_{jk} \Delta_{jk}, \label{double}
\end{eqnarray}
with $c_{r} = \sqrt{N} c_{j}$ \cite{saffman}. Elimination of the doubly excited Rydberg state described by Eq. (\ref{double}) by means of an
adiabatic approximation yields
\begin{eqnarray}
\dot{c}_{g} &=& \sqrt{N} \Omega c_{r}, \\
\dot{c}_{r} &=& -\sqrt{N} \Omega c_{g} + \frac{i \bar{\Delta} \Omega^2}{N} c_{r}, \label{solution}
\end{eqnarray}
where $\bar{\Delta} = \sum_{j,k>j}^{N} \frac{1}{\Delta_{jk}}$ is the mean dipole shift. The solution of Eq. (\ref{solution}) for $c_{g}(0) = 1$
(initially all atoms are in their ground state $| g \rangle$) reads as
\begin{equation}
|c_{r}(t)|^2 = \mbox{sin}^2(\sqrt{N l} \Omega t)/l,
\end{equation}
with $l = 1 + \frac{\bar{\Delta}^2 \Omega^2}{4 N^3}$. The evolution from collective state $|e \rangle$ to
singly excited state $| r \rangle$ in time $t = \frac{\pi}{2 \sqrt{N l} \Omega}$ occurs with probability $P_{1} = 1/\l$.
In the limit of finite dipole blockade, the probability of unwanted double excitations after the $\pi$ pulse is given by
\begin{equation}
P_{2} = \sum_{j,k>j}^{N} |c_{jk}|^2 = \frac{\bar{\Delta}_{P_{2}} \Omega^2}{N}, \label{P2}
\end{equation}
with $\bar{\Delta}_{P_{2}}
 = \sum_{j,k>j}^{N} \frac{1}{\Delta_{jk}^{2}}$. A finite blockade also implies a frequency shift of the effective two-level system ($|g\rangle$ and $|r\rangle$).
The resonance frequency is shifted by $\delta \omega = \frac{\Omega^2 \bar{\Delta}}{N}$.

The above results can be applied to the case of any single-qubit operations. We assume that a qubit is realized
by a quasi one-dimensional (cigar shaped) atomic vapour consisting of $\sim$ 300 $^{87}$Rb atoms.
The spatial distribution (probability density) of an atomic cloud is given by
\begin{equation}
P(z) = (2 \pi \sigma^2)^{-1/2} \mbox{exp}(-z^2/2 \sigma^2) \label{distribution},
\end{equation}
where $z$ is a dimension along the ensemble and $\sigma = 3.0$ $\mu\mbox{m}$ is the variance.
The level $|r\rangle$ may correspond to $43D_{5/2}$ or $58D_{3/2}$. The probability of double excitation given by Eq. (\ref{P2})
can be rewritten in terms of the mean blockade shift $B$ \cite{walkerbloc}. Hence, the probability of double excitation is
$P_{2} =  \Omega^2_{N} (N-1)/2 N B^2$, where $\Omega_{N} = \sqrt{N} \Omega$. For $43D_{5/2}$ and $58D_{3/2}$, the mean
blockade shift is $B = 0.25 \ \mbox{MHz}$ and $B = 2.9 \ \mbox{MHz}$ in a
trap with $\sigma = 3.0$  $\mu\mbox{m}$, respectively \cite{walkerbloc}. Hence, the probability of double excitation for the $43D_{5/2}$ level
is $P_{2} \cong 2.3 \ 10^{-3}$ and for the $58D_{3/2}$ level is $P_{2} \cong  1.7 \ 10^{-5}$.

The fidelity of the single-qubit rotations can be as high as $F_{single} = \exp(- 2 P_{2}) = 0.999$, where $P_{2} = 1.7 \ 10^{-5}$
(the probability of doubly-excited states and singly-excited states outside the desired two-level system are similar).
Above fidelity is given for the worst case scenario when the separation of
atoms is maximal and the dipole-dipole interaction is of the weakest (van der Waals) type.
The time of a $\pi$-pulse applied to the transition between $|e \rangle$ and $| r \rangle$ is $\sim$ 14.6 $\mu$s. We estimate
that the rest of the $\pi$-pulses which are necessary to realize any single-qubit rotation (see Fig.~\ref{X&H}) can be applied in time significantly shorter than the above time.

In the above experimental implementation, the single-qubit rotations can be carried out on a microsecond timescale. Hence, the spontaneous emission
from the Rydberg state and the black-body transfer (to other Rydberg states) which occur with low rates of order $10^{3}$ Hz are negligible. Moreover, a low temperature of the atomic vapour implies low atomic collision rate and negligible Doppler broadening.

The single-qubit rotations are one of the basic operations that are necessary in any model of quantum computation.
The above fast and reliable implementations of the single-qubit operations open a possibility for a realization of the measurement-based model
of quantum computation. However, we are still lacking a scheme for efficient generation of the cluster states.
Before introducing an efficient protocol for cluster states generation, let us review the concept of the measurement-based model
of quantum computation realized on the cluster states.

\section{Entangling operations for the creation of cluster states}
\subsection{Cluster states and one-way model of quantum computation}
There are many approaches to scalable quantum computing (QC), such as the standard circuit model of QC, adiabatic QC
or measurement-based QC, realized on the cluster states \cite{koklecture,nielsen_cluster}.
Cluster states are large entangled states that act as a universal resource for the one--way quantum
computer \cite{rauss,hein,koklecture}. Graphically cluster states are represented in a form of lattice or a graph. We associate with every
node $j$ of a graph an isolated qubit in the state $| + \rangle = \frac{1}{\sqrt{2}} (|0\rangle_{j} + |1\rangle_{j})$  connected (entangled) with adjacent qubits via the $CZ_{jk}$ (controlled-Z) operations
\begin{equation}
CZ_{jk} = |0\rangle_{j}\langle0| \otimes \mathbb{\hat{I}}_{k} + |1\rangle_{j}\langle1| \otimes Z_{k},
\end{equation}
where $|0\rangle$, $|1\rangle$ are the computational basis states and $Z$ is the Pauli $\sigma_{z}$ operator.
Commonly, cluster states are described in terms of the stabilizer operators.
A set of commuting operators $S_{j}$ constitutes a stabilizer of quantum state $|\phi\rangle$ under which the state is invariant.
The stabilizer formalism allows us to describe quantum states of a set of qubits and its evolution in terms of few stabilizer operators,
which usually consist of operators from the Pauli group $G_{n}$.
The state $|\phi_{N}^{C}\rangle$ of cluster $C$ consisting of $N$ qubits
is completely specified by the following set of eigenvalue equations:
\begin{equation}
S_{j} |\phi_{N}^{C}\rangle=|\phi_{N}^{C}\rangle,
\end{equation}
with
\begin{equation}
S_{j} = X_{j} \prod_{k \ \in \ \mbox{nghb}(j)} Z_{k},
\end{equation}
where $\mbox{nghb}(j)$ is the set of all
neighbours of qubit $j$ \cite{rauss}. The $S_{j}$  are Hermitian stabilizer operators whose eigenstates (the cluster states) are mutually orthogonal and
form a basis in the Hilbert space of the cluster \cite{rauss}.

\begin{figure}[!h]
\begin{center}
\includegraphics[scale=1]{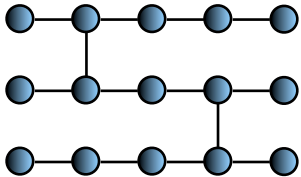}
\end{center}
\caption{A cluster state. Nodes represent physical qubits which are connected via the $CZ$ operations. A horizontal strings of
physical qubits constitute a logical qubit. The vertical links between logical qubits represent two-qubit $CZ$ gates.\label{graph}}
\end{figure}

In the measurement-based model of QC, the entire resource for quantum computation is provided from the beginning as a cluster state [Fig. \ref{graph}].
Quantum computation consists of single-qubit measurements on the cluster states and every quantum algorithm is encoded in a measurement blueprint.
Measurement of a qubit in the $Z$ eigenbasis (in the computational basis) removes a qubit from a cluster and all links to its neighbours are broken.
In conclusion, a cluster is reduced by one qubit, and possible corrective $Z$ operation is applied to its neighbours depending on the measurement output
(if the measurement result is $|0\rangle$ then noting happens, but when the measurement output is $|1\rangle$ a phase-flip is applied). By means of a
$Z$ measurement, any cluster can be carved out from a generic, fully connected cluster (Fig. \ref{graph}).
Other measurements are performed in the basis $B(\alpha) \in \{|\alpha_{+}\rangle,|\alpha_{-}\rangle\}$ where
$|\alpha_{\pm}\rangle = \frac{1}{\sqrt{2}} (e^{i \alpha/2}|0\rangle \pm e^{-i \alpha/2}|1\rangle)$. For $\alpha = 0$ the measurement is realized
in the $X$ eigenbasis. An interesting feature of $X$ measurement is that two neighboring $X$ measurements in a
linear cluster remove measured qubits and connects their neighbours with each other resulting in a shortened cluster.
For $\alpha = \pi/2$, the $Y$ measurement is performed. In result of the $Y$ measurement,  the measured qubit is removed from a cluster and
its neighbours are connected (up to a corrective phase operation).
Measurements in the $X$ and $Y$ eigenbases propagate quantum information through cluster. In general, any quantum computation proceeds as a series of
measurement governed by an appropriate blueprint. The choice of measurement basis for every physical qubit is encoded in this measurement blueprint.
Moreover, all measurement bases depend on the outcomes of the preceding measurements. This is called feed-forward operation.
Although the result of any measurement is completely random, information processing is possible because of the feed-forward operations. The feed-forward
operations ensure that measurement bases are correlated and deterministic computation can be realized. In this way quantum information propagates
through cluster until the last column of qubits, which are then ready to be read out. Readouts are performed in the
$Z$ eigenbasis up to Pauli correction and the output of the computation is a classical bit string \cite{rauss}.

Cluster states are a promising resource for quantum information processing and quantum computer.
One possible way of creating arrays of qubits is trapping single atoms or small atomic ensembles in optical lattices.
However, since a cluster consists of a large set of entangled qubits, efficient protocols for
generating entanglement between pairs of qubits in lattice are required.

\begin{figure}
\includegraphics[scale=0.60]{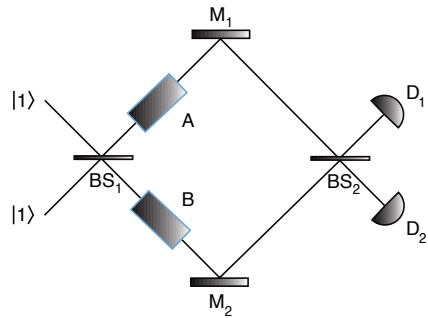}
\caption{Diagram of the protocol. We send an entangled pair of photons in
the state $|\phi_{light}\rangle=\frac{i}{\sqrt{2}} (|02\rangle +
|20\rangle)$ into the arms of a Mach-Zehnder interferometer. The photons interact with atomic
vapours: One and only one alkali atom in the ensemble is excited by one
of the photons to the Rydberg state $|r\rangle$. Absorption of the
second photon is prohibited by the dipole blockade mechanism. After $BS_{2}$,
the state of ensemble-light system has the following form $|\phi_{out}\rangle=\frac{i}{\sqrt{2}}
(|\psi^{+}\rangle |01\rangle + |\psi^{-}\rangle |10\rangle)$, where
$|\psi^{\pm}\rangle = \frac{1}{\sqrt{2}}(|re\rangle \pm i
|er\rangle)$. Detection of a single photon will leave the atomic ensembles
entangled. \label{scheme}}
\end{figure}

\subsection{Protocol - entangling operation}
We propose a scheme for cluster state generation, based on the dipole blockade mechanism.
The entangling operation is constructed as follows: Two atomic ensembles
are placed in the arms of a Mach-Zehnder interferometer (see
Fig.~\ref{scheme}). Initially, we prepare each ensemble $A$ and $B$ in
the state $|\phi_{A,B}\rangle = |e\rangle \equiv |e_{1},e_{2}, \ldots
,e_{N}\rangle$ (see Fig.~\ref{levels}). Two indistinguishable photons enter each input mode of
the interferometer, and due to the Hong-Ou-Mandel (HOM) effect, after
the first beam splitter ($BS_{1}$), the two photons are in the state
$|\phi_{light}\rangle=|11\rangle \xrightarrow{BS_{1}} \frac{i}{\sqrt{2}}
(|02\rangle + |20\rangle)$ where $|0\rangle$ and $|2\rangle$ denote the
vacuum and a two-photon state, respectively. After $BS_{1}$ the photons
interact with the atomic ensembles: One and only one atom in the
ensemble is excited by one of the photons to the Rydberg state
$|r\rangle$, and the absorption of the second photon is prohibited
by the dipole blockade mechanism. The total state of a ensemble-light
system after interaction is given by
\begin{equation}
|\phi_{int}\rangle=\frac{i}{\sqrt{2}} (|er\rangle |01\rangle +
|re\rangle |10\rangle).
\end{equation}
After the second beam splitter ($BS_{2}$), the total state is
\begin{equation}
|\phi_{out}\rangle=\frac{i}{\sqrt{2}} (|\psi^{+}\rangle |01\rangle +
|\psi^{-}\rangle |10\rangle),
\end{equation}
where $|\psi^{\pm}\rangle = \frac{1}{\sqrt{2}}(|re\rangle \pm i
|er\rangle)$. Conditional on a single photodetector click, the
ensembles are projected onto a maximally entangled state. After
establishing entanglement, the qubits are transferred to their
computational basis states $|0\rangle_{L} \equiv |g\rangle$ and
$|1\rangle_{L} \equiv |s\rangle$ by classical optical pulses
$\Omega_{g}$ and $\Omega_{s}$. This means that ideally every run of the
procedure would give an entangled state of ensembles with success
probability $p_{success} = \eta$, where $\eta$ is the combined detection and source
efficiency. This is a significant improvement compared to the
success probability $p_{success} = \eta^2/2$ of the double heralding protocol in
Ref.~\cite{kok}.

\begin{figure}
\includegraphics[scale=0.60]{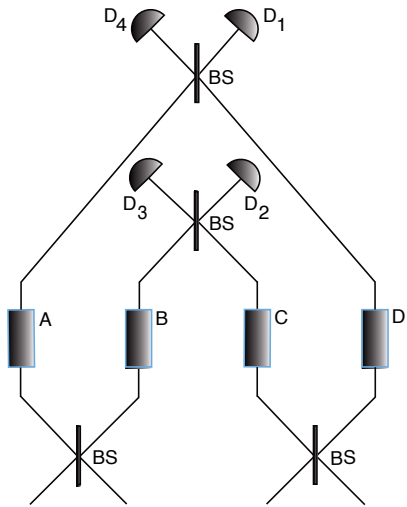}
\caption{The scheme for creating the 4-qubit GHZ state. The four ensembles
$A$, $B$, $C$, and $D$ are prepared in the state
$|\phi_{ABCD}\rangle=|eeee\rangle$. Four indistinguishable photons are sent into
the beam splitters. The interaction of photons with the atomic
vapours is followed by the beam splitters and four photodetectors.
Conditional on photodetector clicks at the photodetector $(D_{1},D_{2})$,
$(D_{1},D_{3})$, $(D_{4},D_{2})$ or $(D_{4}, D_{3})$,
a state of the four qubits is projected onto the 4-qubit GHZ
state (up to phase correcting operations) with success probability $p_{success} =
\eta^2/2$. \label{GHZ}}
\end{figure}

\subsection{Generation of the GHZ and cluster states}
The entangling operation can be used to efficiently create arbitrary
cluster states, including universal resource states for quantum
computing. However, a modification of the entangling procedure yields an
even more dramatic improvement in the efficiency of cluster state
generation. By arranging the ensembles in a four-mode interferometer as
shown in Fig.~\ref{GHZ}, the detection of two photons will create the
four-qubit GHZ state \emph{in a single step}. Moreover, since only two photons
are detected, the protocol is relatively insensitive to detector losses.
The success probability is $p_{success} = \eta^2/2$. Higher GHZ states can
be created by a straightforward extension. A subsequent cluster states
are generated with success probability
\begin{equation}
p_{success} = \eta^{Q/2} (Q-2)/2^{Q-2},
\end{equation}
where $Q = 4, 6, \ldots$ is the number of the qubits.

The efficiently generated large GHZ states may serve as building blocks
for cluster states. By entangling small clusters with
the above entangling procedure, large cluster states can be constructed.
A single photon applied to a pair of qubits (each from two different
4-qubit cluster states) followed by a single photodetector click creates
a 8-qubit cluster state with success probability $p_{success} = \eta'/8$. This
procedure can be repeated in an efficient manner \cite{kieling}. In case
of failure, the two qubits that participated in linking are measured in
the computational basis, and the rest of the cluster state is recycled \cite{nielsen_microcluster}.

\section{Errors, decoherence mechanisms and fidelity}
The dominant errors and decoherence mechanisms that enter the entangling operation are the following:
\emph{(i)} The coincident event in the HOM effect,
\emph{(ii)} the spontaneous emission rate of the Rydberg state,
\emph{(iii)} the black-body transfer rate (to other Rydberg states),
\emph{(iv)} the atomic collision rate,
\emph{(v)} the doubly-excited Rydberg states and singly-excited states outside the desired two-level system,
\emph{(vi)} no absorption event, and
\emph{(vii)} the inefficiency and the dark count rate of the photodetectors.
We analyze in more detail the dominant error and decoherence mechanisms on the following experimental implementation.

\subsection{Experimental implementation of the entangling operation}
\begin{figure}
\includegraphics[scale=0.60]{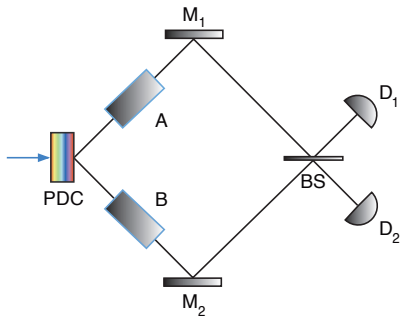}
\caption{Example of an experimental implementation of the protocol. The source of a single-photon pair consists of the type I nonlinear crystal.
Generated single-photon pulses are entangled in the momentum (path) degree of freedom. \label{implement}}
\end{figure}

First, consider the coincident events in the HOM effect. The single
indistinguishable photons that recombine at the first beam splitter
($BS_{1}$) can be generated by means of the spontaneous parametric
down-conversion (SPDC) process or any other source of a single-photon pulses such as atomic ensemble inside an optical cavity \cite{thompson}.
However, we believe that the SPDC source is better suited for the entangling protocol because of a very complex setup structure of the atomic-based source. The SPDC source (the non-linear crystal) must be pumped with a narrowband ($\sim$ 1 MHz) laser or placed inside a cavity. These kind of cavity-enhanced SPDC sources produce pairs of a single, identical photons with a narrow bandwidth of order of MHz and a spectral brightness of $\sim$ 1500 photons/s per MHZ bandwidth \cite{neergaard,bao}.\\
In general, successful generation of the
entangled state of light depends on the proper setup, where both photons from the SPDC source
recombine at $BS_{1}$ at the same time. In a recent experiment, the coincident event in the HOM effect happens with a
rate of 1500 counts/s \cite{kim}. We assume that the rate of coincident
events is negligible comparing to the time scale of the protocol which
is $t \cong 3$ $\mu$s (the origin of this value will be given later on). In fact, it is possible to completely eliminate the coincident event in the HOM
effect by getting rid of the $BS_{1}$. In place of single-photon sources and $BS_{1}$, one can use a SPDC source generating pairs of
single-photons entangled in momentum (path) degree of freedom \cite{barbieri,rossi}. The state of the photons is given by
$|\phi_{light}\rangle= \frac{1}{\sqrt{2}}
(|20\rangle_{AB} + |02\rangle_{AB})$, where $A$ and $B$ are the single-photon pairs that interact with an ensemble A and B, respectively (see Fig. \ref{implement}).
Moreover, since the SPDC process is a phase and energy matching phenomenon, no phase difference appears between two paths (pairs) $A$ and $B$   \cite{barbieri}.
In general, the whole Mach-Zehnder interferometer
needs to be phase-stable. In case of a GHZ state generation, phase locking of a large number of Mach-Zehnder interferometers is very demanding
(although possible). Therefore, by replacing single-photon sources and $BS_{1}$ with the SPDC source generating entangled photon pairs we
solve another potential roadblock to an experimental realization of the entangling operation.
Recently, it has been shown that these kind of entangle pairs of photons can be generated very effectively \cite{rossi}.

Now, assume that an atomic vapour consists of 300 $^{87}$Rb atoms placed in
the far off-resonant optical trap or magneto-optical trap (MOT). The
atomic levels $|g \rangle$, $|e \rangle$, and $|r \rangle$ may correspond
to $(5S_{1/2}, F = 1)$, $(5P_{3/2}, F = 2)$ and $43D_{5/2}$ or $58D_{3/2}$, respectively.
State  $|s \rangle$ may correspond to the hyperfine state $(5S_{1/2}, F = 2)$, which implies that the transition from $|s\rangle$ to $|r\rangle$
is a two-photon process (see Fig.~\ref{levels}). We have identified state $|e\rangle$ with a short lived state $(5P_{3/2}, F = 2)$, when in fact it must be a
long lived energy level. However, in case of the MOT trap, the requirement of a long relaxation time of the state  $|e \rangle$ can be lifted
since the trap lasers produce a constant population in the $|e \rangle$ state \cite{li,monroe}. In general, a requirement of state $|e \rangle$
is imposed to simplify experimental realization of the protocol where usually two--photon excitation are used to obtain Rydberg atoms.
The spatial distribution of an atomic cloud is a quasi one-dimensional (cigar shaped)
ensemble with probability density given by Eq. (\ref{distribution}). Atomic
vapours described with quasi  one-dimensional probability density have
been demonstrated experimentally \cite{johnson}.\\
When a protocol is based on a quantum optical system, its performance is
limited by the inefficiency and the dark count rate of the
photodetectors. The dark count rate of the modern photodetector $\gamma_{dc}$ can be
as low as 20 Hz and efficiency reaches $\eta \approx 30\%$ for
wavelengths around 480 nm. The probability of the dark count is
$P_{dc} = 1 - \mbox{exp}(-\gamma_{dc} t/p_{success} ) \cong 3.2 \ 10^{-4}$.
In general, the probability of the dark count is negligible for $p_{success} > \gamma_{dc} t$.\\
Since the length of the atomic ensemble needs to be of an order of several $\mu$m,
the most important source of errors is the lack of absorption event. The
probability of an absorption of a single photon by a cigar shaped atomic ensemble is
given by $P_{abs} \cong  1 - e^{-N_{i} \sigma_{0} /A}$, with $N_{i}$ the
number of atoms in the interaction region, $\sigma_{0} = 3 \lambda^2/(2
\pi)$ is the on-resonance scattering cross section of a single-photon pulse, and $A = \pi w^2_{0}$ is the area of a single-photon pulse
with a waist $w_{0}$ \cite{walkerabs}. With  $\lambda_{43D} =
485.766$ nm, $\lambda_{58D} = 485.081$ nm, and $w_{0} \approx \pi
\lambda$, the probability of an absorption for both Rydberg states is
$P_{abs} \cong 0.989$.\\
The probability of doubly-excited Rydberg states (absorption of both
photons by an ensemble) depends on the quality of the dipole blockade
and was given  in Sec. III. Here, we rewrite it in terms of the mean
blockade shift $B$ \cite{walkerbloc}. Hence, the probability of doubly-excited Rydberg states is
$P_{2} = g^2_{N} (N-1)/2 N B^2$, where $g_{N} = \sqrt{N} g_{0}$ with $g_{0}$ the atom-light coupling constant written as
\begin{equation}
g_{0} = \frac{\sqrt{2} \rho}{\hbar} \sqrt{\frac{\hbar \omega}{2 \epsilon_{0} V}},
\end{equation}
with $\rho$ being the dipole moment of the $|e\rangle$-$|r\rangle$ transition, $\omega$ being the frequency of the optical field and
$V$ being the quantization volume of the two-photon radiation field.
For $43D_{5/2}$ and $58D_{3/2}$, the mean
blockade shift is $B = 0.25 \ \mbox{MHz}$ and $B = 2.9 \ \mbox{MHz}$ in a
trap with $\sigma = 3.0$  $\mu\mbox{m}$, respectively \cite{walkerbloc}.
Hence, the probability of doubly-excited states for the $43D_{5/2}$ level
is $P_{2} \cong 0.26 $ and for the $58D_{3/2}$ level $P_{2} \cong 0.57 \ 10^{-3}$. The probability of doubly-excited states and
singly-excited states outside the desired two-level system are similar.
The above probabilities are given for the worst case scenario when the separation of
atoms is maximal and the dipole-dipole interaction is of the weakest (van der Waals) type.\\
The time scale of the entangling protocol $t \cong 3$ $\mu$s consists of a time given in Sec. III with $\Omega \equiv g_{0}$ and time of a $\pi$-pulse with the Rabi frequency $\Omega_{s} \cong 1$ MHz applied after a single
photodetector click.\\
The spontaneous emission from the Rydberg state and the black-body
transfer (to other Rydberg states) occur with rates of order $10^3$ Hz,
and are negligible, since after successful entanglement preparation the state
of ensemble is stored in the long lived atomic states  $|g \rangle$ and
$|s \rangle$. Exact values of these rates are given in Refs.
\cite{kim,day}. The atomic collision rate is given by
\begin{equation}
\tau^{-1}_{col}\approx n \sigma_{col}/\sqrt{M/3 k_{B} T},
\end{equation}
with $n$ the number density of atoms, $\sigma_{col}$ the collisional cross section
($\sim 10^{-14}$ $\mbox{cm}^2$), $M$ the atomic mass, $k_{B}$ the
Boltzmann's constant, and $T$ the temperature  \cite{james}. Assuming a
vapour with the number density of atoms of order $10^{12}$  cm$^{-3}$
and the temperature of $\sim 10^{-3}$ K, the atomic collision rate can be as
low as 2 Hz. Moreover, with a sufficiently large energy difference
between states $|g\rangle$ and $|s\rangle$ a single collision is not
likely to affect the qubit.

A low temperature of an atomic vapour implies negligible Doppler broadening. The Doppler broadening is described by the Gaussian
distribution with a standard deviation of $\Delta \lambda = \lambda_{0} \sqrt{k_{B} T/M c^2}$ where $\lambda_{0}$ is
the center wavelength of the Doppler profile (wavelength of a transition between states $|e\rangle$ and $|r\rangle$).
For both $\lambda_{0}=\lambda_{43D}$ and $\lambda_{0}=\lambda_{58D}$, the Doppler broadening is $\Delta \lambda = 0.5 \ 10^{-6}$ nm.
Hence, the Doppler broadening does not affect fidelity of the protocol.

Considering the time scale of the protocol, the entangling procedure is mostly affected by the no
absorption event and inefficiency of the photodetectors. We assume that
the coincident event rate in the HOM effect is negligible and $|r \rangle$ corresponds
 to $58D_{3/2}$ which implies low double excitation probability. In the presence of the above noise and
decoherence mechanisms, the final state of the system conditional on a
single photodetector click is given by
\begin{equation}
\rho_{fin} = (1 - 2\varepsilon) |\psi^{\pm} \rangle \langle \psi^{\pm}| + 2\varepsilon \rho_{noise} + \mathcal{O}(\varepsilon^2),
\end{equation}
where $|\psi^{\pm}\rangle = \frac{1}{\sqrt{2}}(|sg\rangle \pm i
|gs\rangle)$ and $\varepsilon = 1-P_{abs}$ where $P_{abs}$ is the
probability of an absorption of a single photon by an atomic ensemble.
$\rho_{noise}$ denotes the unwanted terms in the state of the two
ensembles. The source efficiency does not affect the fidelity of the final state,
it only lowers the success probability.
After taking into account all dominant error mechanisms, the fidelity of
the prepared entangled state is given by
\begin{equation}
F = \langle \psi^{\pm} | \rho_{fin} |\psi^{\pm} \rangle \cong 0.982,
\end{equation}
which is close to current fault-tolerant thresholds \cite{nielsen}.\\

\section{Conclusions}
We have presented and analyzed a scheme for cluster state generation
based on atomic ensembles and the dipole blockade mechanism. The
protocol consists of single-photon sources, ultra-cold atomic ensembles, and
realistic photodetectors. The protocol generates \emph{in a single step} GHZ state with success
probability $p_{success}\sim \eta^{Q/2}$, where $Q$ is the number of the qubits,
and high fidelity $F \cong 0.982$. The protocol is more efficient than
any previously proposed probabilistic scheme with realistic
photodetectors and sources. In general, number-resolution photodetectors
are not required.
The GHZ states are locally (up to Hadamard operation) equivalent to star-shaped cluster states. We also reviewed and analyzed a
scheme implementing any single-qubit operation on the qubit defined as collective states of mesoscopic ensemble.
The scheme for single-qubit rotations uses classical optical pulses and the dipole blockade mechanism. The experimental implementation may be carried out with high fidelity $F_{single} \cong 0.999$ and on the microsecond timescale with current state-of-the-art experimental setups.\\
The described protocols for single-qubit rotations and entangling operation open a possibility of experimental implementation of the
measurement-based quantum computer based on atomic ensembles.

\begin{acknowledgments}
We thank Charles Adams, Matthew Jones, and Klaus M{\o}lmer for helpful discussions. This
work was supported by the White Rose Foundation.
\end{acknowledgments}

\end{document}